\documentclass[a4paper,11pt]{article} 
\pdfoutput=1 

\usepackage{jcappub}
\makeatletter
\gdef\@fpheader{}
\makeatother

\usepackage{graphicx,latexsym,amsmath,amssymb}

\usepackage{hyperref}
\hypersetup{colorlinks,citecolor=nicegreen,linkcolor=nicered}
\usepackage{color}
\definecolor{red}{rgb}{1,0.1,0.1}
\definecolor{nicered}{rgb}{0.7,0.1,0.1}
\definecolor{nicegreen}{rgb}{0.05,0.35,0.05}

\renewcommand\afterTocSpace{\vskip 4ex} 

\renewcommand\afterEmailSpace{\vskip25pt plus0.2fil minus12pt}
\renewcommand\afterTitleSpace{\vskip25pt plus0.2fil minus12pt}
\renewcommand{\figurename}{\small Figure}
\renewcommand{\tablename}{\small  Table}

\usepackage{txfonts}

\renewcommand\b{\beta}
\newcommand\g{\gamma}
\renewcommand\r{\rho}
\newcommand\s{\sigma}
\newcommand\be{\begin{equation}}
\newcommand\ee{\end{equation}}
\newcommand\bea{\begin{eqnarray}}
\newcommand\eea{\end{eqnarray}}

\newcommand\kpc{\text{kpc}}
\newcommand\pc{\text{pc}}
\newcommand\se{\text{s}}
\newcommand\km{\text{km}}
\newcommand\GeV{\text{GeV}}
\newcommand\cm{\text{cm}}
\newcommand\Ms{\ensuremath{M_\odot}}

\title{\huge The Dark Matter Halo of the  Milky Way, AD 2013}

\author[a]{Fabrizio Nesti,}
\author[b]{Paolo Salucci}
\affiliation[a]{Gran Sasso Science Institute, viale Crispi 7, I-67100 L'Aquila, Italy}
\affiliation[b]{SISSA/ISAS, Via Bonomea 265,  I-34136 Trieste, Italy}

\emailAdd{\tt nesti@aquila.infn.it}
\emailAdd{\tt salucci@sissa.it}


\arxivnumber{1304.5127}
       
\abstract{%
We derive the mass model of the Milky Way (MW), crucial for Dark Matter (DM) direct and indirect
detection, using recent data and a cored dark matter (DM) halo profile, which is favoured by studies
of external galaxies.
The method used consists in fitting a spherically symmetric model of the Galaxy with a Burkert DM
halo profile to available data: MW terminal velocities in the region inside the solar circle,
circular velocity as recently estimated from maser star forming regions at intermediate radii, and
velocity dispersions of stellar halo tracers for the outermost Galactic region.  The latter are
reproduced by integrating the Jeans equation for every modeled mass distribution, and by allowing
for different velocity anisotropies for different tracer populations.
For comparison we also consider a Navarro-Frenk-White profile.  We find that the cored profile is
the preferred one, with a shallow central density of $\rho_H\sim4\times10^{7}\Ms/\kpc^3$ and a large
core radius $R_H\sim10\,\kpc$, as observed in external spirals and in agreement with the mass model
underlying the Universal Rotation Curve of spirals.
We describe also the derived model uncertainties, which are crucially driven by the poorly
constrained velocity dispersion anisotropies of halo tracers.
The emerging cored DM distribution has implications for the DM annihilation angular profile, which
is much less boosted in the Galactic center direction with respect to the case of the standard
$\Lambda$CDM, NFW profile.  Using the derived uncertainties we discuss finally the limitations and
prospects to discriminate between cored and cusped DM profile with a possible observed diffuse DM
annihilation signal.
The present mass model aims to characterize the present-day description of the distribution of
matter in our Galaxy, which is needed to frame current crucial issues of Cosmology, Astrophysics and
Elementary Particles.%
}

\begin{document}

\maketitle

\bigskip

\section{Introduction}

The evidence for the phenomenon explained in term of Dark Matter in Galaxies is outstanding.  In
particular, spiral rotation curves (e.g.~\cite{rubin80,bosma81}) have revealed the presence in these
objects of a dark ``mass component''.  The {\it current} framework includes a central bulge, a
stellar disk and an extended gaseous disk, all embedded in a spherical halo made by dark particles
(of yet) unknown nature.  In external galaxies, the dark-luminous mass distribution has been derived
by means of a) hundreths of individual rotation curves (e.g.~\cite{2008MNRAS.383..297S}) b) 1000
{\it coadded} RCs~\cite{URC2} and c) various measurements of galaxy's gravitational potential (see
e.g.~\cite{2011arXiv1102.1184S}). The result is that DM halos show a shallow central density
profile, moreover, stellar disk radii, halo core and viral radii, gaseous and stellar masses, turn
out to be all related and to lead to an Universal Rotation Curve~\cite{URC2}, likely the final state
of complex physical processes governing the formation of galaxies.

Let us now consider the Milky Way. It is logic to ask: what is its distribution of dark and luminous
matter? Does it conform to that of external galaxies? Moreover, living ``inside'' the object does
help us in better disentangling the different mass components? Also of interest, do we know well
enough how the DM particles are distributed around the Galaxy, in order to predict the signals of
their annihilations to be expected in current experimental surveys?

It is not easy to address these questions and obtain a proper mass model of the Galaxy.  The MW is
an intrinsically complex object: observations are difficult to interpret due to our location within
it.  We cannot measure the MW circular velocity, and we must resort to more indirect kinematical
measurements to infer its gravitational field.  In fact, inside the solar circle (see
e.g.~\cite{McClure07}) we measure the rotating HI disk terminal velocities $V_T$ that relate to the
circular velocity profile $V(r)$ (with $r$ the galactocentric radius) only once one assumes some
values for $R_\odot$, the distance of the Sun from the Galaxy center, and $V_\odot$, the value of
the circular velocity at the Sun's position (see section~\ref{sec:terminal} below).  Both quantities
are presently known within an uncertainty of 10\% (e.g.~\cite{MB09} that triggers a not negligible
uncertainty in the derived circular velocity $V(r)$.  Outside the solar circle, the circular
velocities $V(r)$ are inferred by applying the Jeans equation to the kinematics of populations of
tracer stars. Remarkably, the result strongly depends on the (unknown) dynamical state of tracers
and it is also affected by uncertainties in their photometric and kinematical
measurements~\cite{battaglia05,xue08,brown09}.  The mass modeling method features
(see~\cite{caldwell81}, and e.g.~\cite{sofue1}): a) an ad hoc halo density profile b) a number of
assumptions on the dynamical status of the tracers and c) on their density distributions.  All this
adds to the complications due to the fact that the MW rotation curve is quite flat and therefore it
is intrinsically difficult to decompose in its dark and luminous components~\cite{Tonini}.

However, there are today good motivations to attempt deriving a robust and reliable MW mass model,
i.e.: to test claims of Gamma-ray emissions from annihilating DM particles in high density regions
of the Galaxy, to investigate the Nature of the dark particles from the galactic DM density
distribution, to understand how the Galaxy and the Local Group formed.  A novel approach is
warranted since in the past few years, modeling techniques and measurements have progressed.

First, let us note that the mass distribution in the Galaxy is likely to be similar to that of any
other galaxy of the same mass/luminosity/dimensions~\cite{URC2}.  More in detail, the DM halo
density profile in spirals is represented by the URC halo profile~\cite{donato09}.  The different
profiles adopted in several previous works, (NFW and/or pseudo-isothermal halo) are not supported by
present day observations in external galaxies.  In addition, in considering the NFW halo profile one
should confront with the standard results emerging from simulations~\cite{2011ApJ...740..102K}, a
procedure not always followed in previous works.

Second, trigonometric parallaxes and proper motions of sources of maser emission associated with
high-mass star forming regions in the disk of the Galaxy have been recently
available~\cite{Brunthaler:2011sy}. These measurements allows one to determine source distances and
proper motions in a direct and geometrical way, and to obtain their full 3-dimensional locations and
velocity vectors leading, for particular line of sights, to the actual Galaxy circular velocity.
This provides us with a (limited) number of true and direct determinations of the velocity speed on
the Galactic disk.

Third, a very extended and large amount of radial motions of stars, from which to extract the
dispersion velocity, has now become available out to 80 kpc and more~\cite{xue08, Gnedin:2010fv,
  Deason:2012wm}.
 
Finally, with a profile independent method, we derived a most conservative value of the DM density
at the Sun $\rho_{DM}(R_\odot)=0.43 \pm 0.11\pm0.11$~\cite{nostro}. This measure will be used as a
cross-check of the Galaxy modeling.
 
The aim of this work is to obtain, despite various uncertainties and difficulties still present, a
robust MW mass model of the Galaxy, at the best of the knowledge of 2013.  In the next section we
will present the adopted mass components, in section~\ref{sec:data} we describe the observational
data, in section~\ref{sec:fit} we report the fits to these data and their results; in
sections~\ref{sec:direct},~\ref{sec:ann} we describe the consequences for direct and indirect DM
detection, and in ~\ref{sec:conclusions} we discuss our conclusions.

\section{Mass Components}
\label{sec:components}

In modeling the Galaxy, the main mass components are: the central bulge, the (thin) stellar disk and the
DM halo.  The observational uncertainties and the systematic biases on these components, which will
be discussed below, make negligible (for our aims) other minor mass components present: the central
bar, the thick and HI disks, and the stellar halo.

\paragraph{Bulge.} The MW central bulge region has been subject to a number of probes which along
the years have increased our knowledge of its mass distribution. Nevertheless, due to severe
extinction in the very central region in diverse wavelengths, there remain crucial uncertainties in
the inner 1--2\,\kpc. In particular, the state of the art~\cite{PR04,bissantz04,robin12} knowledge
seems to point quite clearly to a two-components structure: a more massive `boxy' bulge extending up
to $\sim 1.5\,\kpc$ superimposed on a minor subleading `bar', extending up to $\sim 4\,\kpc$.  The
estimates of the total mass of these components is however very difficult, because observations can
only probe the mass density at a minimal angle of $2^\circ$ from the galactic plane, i.e.\ is at the
outer border of the boxy bulge distribution, thus requiring some sort of arbitrary modeling of the
distribution in order to infer the total mass. Kinematical probes such as surveys of the rotational
velocities are also limited by the large velocity anisotropies (see e.g.~\cite{soto12}).  As a
result, the total bulge mass has to be considered as an unknown parameter, subject to a lower bound
which we take conservatively as $\sim 0.8\text{--}1\times10^{10}\,\Ms$.

Due to the above uncertainties and non circularities evident in the central region we will restrict
our analysis of the rotation curve to radii larger than $2.5\,\kpc$. At these distances, the shape
of the mass distribution of the (dominant) bulge component is irrelevant, and the bulge can be
safely be treated as a point-mass in the center. No results in this paper depend on the bulge
density profile, as opposed to the value of its mass that plays a crucial role in mass modeling.

\paragraph{Disk.} As in most of Spirals the MW surface density of the stellar disk is exponential,
which for the thin disk we parametrize as~\cite{Freeman}
\be
\s_D=\frac{M_D}{2\pi R_D^2}{\rm e}^{-r/R_D}.
\ee
with $R_D=2.6\pm0.5\,$kpc~\cite{robin08}.  

The value of the thin disk mass $M_D$ is rather unknown, although it can be constrained to lie in
the range $M_D=5\text{--}7 \times10^{10}\,\Ms$~\cite{PR04,juric08,robin08,reyle09}.  At the same
time the estimate and uncertainty of the disk scale length $R_D$ may be subject to significant
biases (substructures, coverage, etc, see~\cite{robin08}).  The disk thickness is constrained to be
of the order of few hundreths of parsecs and in this work its presence and its variations have a
negligible impact on the stellar contribution to the MW rotation curve (less than few \%, see
e.g.~\cite{nostro}).  It is therefore justified to take the infinitesimal thin disk approximation.
At the same time, the thick and HI disks, whose masses are estimated to be $\sim$ one tenth of the
thin disk one~\cite{robin08,nakanishisofue} can be neglected in the present analysis and be
considered to be part of the above range of $M_D$.

\paragraph{Dark Halo.}  We adopt, after~\cite{URC2} and~\cite{donato09}, the Universal Rotation
Curve (Burkert) profile.  In addition, as a comparison, we consider also a NFW dark halo. These
density profiles are parametrized through the density scale $\r_H$ and scale radius $R_H$
\bea
\r_{NFW}&=&\r_H\frac{1}{x(1+x)^2},\\ 
\r_{Bur}&=&\r_H\frac{1}{(1+x)(1+x^2)},
\eea
where $x=r/R_H$.  Notice that $R_H$ has a different physical significance in the two models: the
radius at which $d\log \r_{NFW}/d\log r= -2$ and the radius of the region of approximately constant
density.

\section{Observational Data}
\label{sec:data}

\subsection{Terminal velocities}
\label{sec:terminal}

The circular velocities inside the solar circle $R_\odot$ can be ``reconstructed'' from HI terminal
velocities $V_{Ti}$ as $(r_i,V_i)=(r_i, V_{Ti}+ y_i V_\odot)$, where $y_i=r_i/R_\odot$ and $r_i$ is
the galactocentric radius of the individual measured gas clouds.  Notice that $y_i$ is actually an
angular measurement, with uncertainty which is negligible for the present work.  The fit consists in
comparing $V_{Ti}$ with the terminal velocities as predicted by the model. In doing so, $y_i$, is
translated into $r_i$, thus involving $R_\odot$ in addition to $V_\odot$.

\begin{figure}[t]
\centerline{\includegraphics[width=.55\columnwidth]{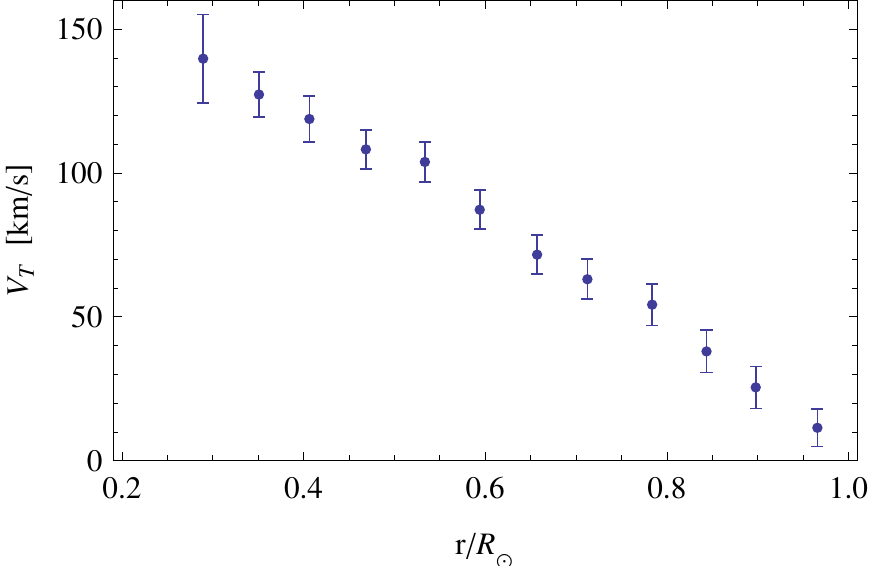}~~~~~~~}
\vspace*{-1ex}
\caption{Terminal velocities, rebinned, see appendix~\ref{app:terminali}.\label{fig:VT}}
\end{figure}

Data from a number of works~\cite{malhotra95,clemens85,alvarez90,McClure07} are reported in
appendix~\ref{app:terminali} and are binned in intervals of $\sim 0.5\,\kpc$, see
figure~\ref{fig:VT}.  The binning procedure is delicate and we stress that some care has to be used
in estimating the uncertainties later used for the fit.  In fact, the errors on single measures are
very small, even below 1\,km/s, but data themselves are widely spread, due to differences between
the values coming from different regions (e.g.\ at opposite longitudes). As a result, the
uncertainties are mainly data-driven, and they have to be estimated accordingly.  We believe that
binning data separately for each measurements series provides a better description of the
observative constraints with respect to the usual procedure of a global increase of the individual
errors. We describe the binning in appendix~\ref{app:terminali}.

\subsection{Maser velocities } 
   
Very precise measures of position and proper motions of maser sources, located at diverse
galactocentric distances, e.g.~\cite{reid09, sanna, Honma:2012zd} provide us with a number of
circular velocities.  While the galactocentric radii referred to these measurements are not crucial
(in view of the almost flatness of the MW rotation profile) their amplitude is important in that it
can resolve the degeneracy that $V_\odot $ and $ R_\odot$ have.  The outermost of such measurements
reach the galactocentric radii of $\sim 13\,\kpc$ and hint to a flat rotation curve, $V(r\simeq
13\kpc)\simeq V(R_\odot)$.  They also set the scale of the circular velocity, otherwise uncertain
between 200\,\km/\se\ and 250\,\km/\se, to the upper end, $V_\odot\sim 250\,\km/\se$

Thus, the two important outcomes of the maser observations~\cite{bovy,Brunthaler:2011sy} are: i) the
quantity $\omega=V_\odot/R_\odot= 30.3\pm 0.9$, consistent with the previous determinations through
the motion of Sagittarius A$^*$, and ii) a direct estimate of the sun circular velocity of
$V_\odot\simeq 239\pm 7$km/s, which is also in agreement with the knowledge of the Sun's
galactocentric radius via $\omega$~(see e.g.~\cite{Schonrich:2012qz}). See also~\cite{Honma:2012zd}
for more recent similar measurements.

These two results play an important role in the mass modeling: we will use i) as a constraint
between $V_\odot$ and $R_\odot$ because its precision is much higher than the uncertainties in other
quantities, and use ii) as the crucial measure of the Sun's LSR rotation velocity.

\subsection{Stellar halo velocity dispersions}
\label{sec:dispersions}

From the measured radial velocities, the velocity dispersion $\s_i^2(r)$ of different halo
populations of tracer (bright) stars have been estimated from the solar neighborhood out to
$\sim80\,\kpc$. They provide unique constraints on the rotation curve at large radii and thus on the
DM density profile.  In particular, as we will see below, the halo scale radius $R_H$ is
significantly related to these measurements and to their uncertainties, that thus require a detailed
discussion.

Assuming virialization, each population traces the gravitational potential, and we can use the
spherical Jeans equation to link the measured velocity dispersion and the Galaxy gravitational
potential. The density of each population is well represented by power law $\r_i\propto r^{-\g_i}$,
so that the Jeans equation can be written as
\be
\label{eq:jeans}
V^2= \s_i^2 \left[\gamma_i-2\b_i -\frac{\partial \ln\s_i^2}{\partial \ln r}\right].
\ee
Note that we must allow different $\b_i$ for each population of tracers, because although they share
the same gravitational potential there is no argument for them to have the same kinematics, $\b_i$
and $\s_i$.

\begin{figure}
\centerline{\includegraphics[width=.55\columnwidth]{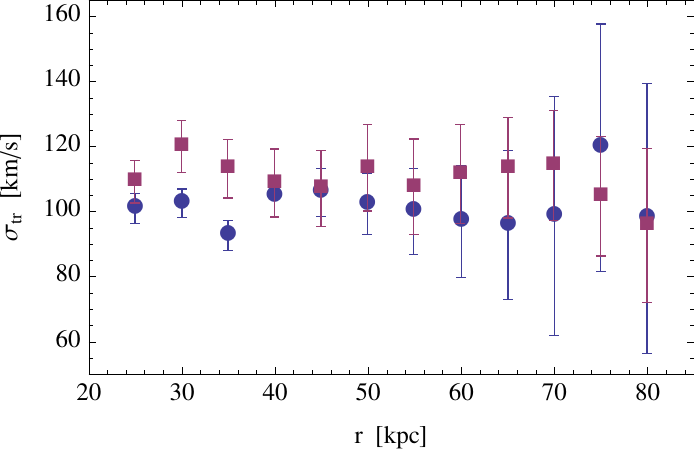}~~~~~}
\vspace*{-1ex}
\caption{Tracers velocity dispersions: ``1'' Gnedin~\cite{Gnedin:2010fv} (red squares), and ``2''
  Xue~\cite{xue08} (blue circles).\label{fig:dispersions}}
\end{figure}

The Jeans equation~(\ref{eq:jeans}) relates the circular velocity model $V^2(r)$, the velocity
dispersions $\s_i(r)$ and the anisotropies $\b_i$. Thus, for each given velocity model $V(r)$ the
equation can be integrated to find the predicted velocity dispersion as a function of radius
$\s_i(r)$.  For each population separately, the dispersion profile so computed from the model can
then be directly compared with data.  The integrated $\s_i(s)$ is determined up to a free constant
multiplied by a known function, solution of the homogeneous equation, $\s^2_{hom,i}\propto{\rm
  e}^{\int(\g_i-2\b_i)\,{\rm d}\ln r}$.  Considering for instance a linear dependence of the
anisotropy on the radius, $\b_i(r)=\b_i+\b'_i r$, the function is simple, $\s^2_{hom,i}=\bar\s_i^2
\,r^{\g_i-2\b_i} {\rm e}^{-2\b_i' r}$, where $\bar\s_i$ is the free constant, that we define so that
$\s_i(80\,\kpc)=\bar\s_i$.\footnote{We prefer to state explicitly this boundary condition at finite
  radius, rather than considering vanishing $\s_i$ at infinity, which is not observationally
  motivated and may well turn out to be untrue.}  Although the constants $\bar \s_i$ enter the fit
as additional free parameters of the mass model, their effect is actually very limited, because the
dispersion kinematics is very uncertain at large radii (see figure~\ref{fig:dispersions}). In
practice, the effect of $\bar \s_i$ is just to reduce even more the constraining power of the data
at large radii; So, for simplicity and without modifying the fit results, we set
$\bar\s_{1,2}=105\,\km/\se$.  We checked directly that also the derivative $\b'_i$ of the tracers
dispersion anisotropy has a negligible impact on the fits, due to the large uncertainties of the
observed dispersion velocities and to the requirement that, at large radii, $\b_i(80\,\kpc)$ should
not be unphysical. The dependence on $\beta'_i$ can well be mimicked by a minor shift of the
constant anisotropy $\b_i^0$, and given the uncertain situation with the latter, we can safely set
$\b'_i=0$.

The procedure of integrating the Jeans equation is of course equivalent to reconstructing
\emph{pseudo}-values of $V^2$ from the measures of $\s^2$, which however would lead to increased and
correlated errors due to the derivative in (\ref{eq:jeans}). For this reason we prefer to fit the
actual kinematical data $\s_i^2$.

\medskip

We use two populations, the (mostly) BHB stars with velocities surveyed by
HVS~\cite{Gnedin:2010fv}, hereafter ``1'', and the SDSS DR6 BHB stars~\cite{xue08} hereafter
``2''.  Both series of binned data are taken as recalculated in~\cite{Gnedin:2010fv} and are
displayed in figure~\ref{fig:dispersions}.  For simplicity we consider spherical tracers density
distributions, for which one has $\g_{1}\approx 4$ and $\g_{2}\approx 3.5$. These values provide a
good spherical fit to BHB density and velocity dispersions as confirmed also by a recent detailed
survey, see~\cite{Deason:2012wm}.  A very recent study using again BHB stars from the SDSS
DR8~\cite{Kafle:2012az} confirms values for the radial velocity dispersions practically coincident
with the older analysis of~\cite{xue08} that we use.  This new analysis also provides an estimate of
the dispersion anisotropies at various radii. For $r>16\,\kpc$ it points to an average anisotropy of
$\sim -0.6\pm0.5$.\footnote{In \cite{Deason:2012wm}, the velocity dispersion anisotropy at large
  radii was estimated to be $\b_{2}\sim+0.5$, by using an incorrect lower rotation curve as a prior
  knowledge, in particular with $V_\odot\simeq220$\km/s.} This is obviously an extremely useful
information for determining the outer Galactic rotation curve.

We will not consider in this work the tracers velocity dispersions at small radii $r<25\,\kpc$,
because: i) data show a clear break both in the density and in the velocity
dispersion~\cite{Deason:2012wm}; ii) the dispersion anisotropy seems to vary rapidly in that region;
iii) at such small radii the effect of the updated values of $V_\odot$ and $R_\odot$ may have to be
reconsidered in the derivation of the dispersions from data, and iv) at these radii the masers
information on $V_\odot$ already provides a good constraint on the rotation curve.


From the two sets of data in figure~\ref{fig:dispersions} and the Jeans equation one can immediately
infer some conclusion regarding the $\b_i$.  First, we note that both velocity anisotropies appear
to have a fairly limited slope $|\partial \ln\s_i^2/\partial \ln r|\lesssim0.1$, and their average
values are $\sigma_i\sim 100$--$110$.  In the limit of flat rotation curve and considering the
recent estimate of $V_\odot\simeq 240\,$km/s (see previous section), one readily obtains from the
Jeans equation~(\ref{eq:jeans}) $\beta_i\simeq -0.5$--$0.8$. So, the i.e.\ fairly tangential halo
velocity dispersions.  This agrees well with the evidence of a tangential outer
halo~\cite{Kafle:2012az}.
Actually (see below) the preferred anisotropy would be even more tangential, i.e.\ $\b_2\lesssim
-1$, due to the high $V_\odot$.  In this respect, a decreasing rotation curve in place of a flat one
at large radii can help relieving this tension.

As a result, the value of $R_H$ resulting from the fit will be correlated with $\beta_i$, with
higher $\b_i$ corresponding to smaller $R_H$.  This will turn out to be the dominant source of
uncertainty in the predicted values of $R_H$, as discussed below.  The measure of $\b_2$
in~\cite{Kafle:2012az} is still preliminary; accordingly, we will use $\b_2\simeq-0.5\pm 0.5$ as a
fiducial range, keeping in mind possible updates in the future.

Moreover, we note that the dispersion velocities of population 1 (red squares) lie on average above
those of 2 (blue circles). Because at the same time $\g_{1}\sim 4$ is larger than $\g_{2}\sim 3.5$,
to trace the same gravitational potential i.e.\ to satisfy eq.~(\ref{eq:jeans}) with a common
$V^2(r)$, the only possibility is to have $\b_{1}-\b_{2}\sim 0.5$ or more.  This is formally
possible, but since the two samples are not relative to radically different populations, it is
unlikely that their dispersion anisotropies differ so much.  Actually, this difference may be a
systematic bias, probably linked to the different ways the dispersions are estimated from
observative data.\footnote{Considering for instance the 10-20\% overall change in the velocity
  dispersions resulting from different estimation methods, as reported by~\cite{brown09}.}
Therefore, we prefer to leave some tension in the fit than stretching the anisotropy values, and
adopt $\b_1= \b_2+0.5$.

\section{Fit}
\label{sec:fit}

Each choice of model parameters determines $V(r)$, $V_T(r)$ and by means of the integration
procedure described above, the velocity dispersions $\s^2_i(r)$.  We fit
\begin{itemize}
\item the binned terminal velocities of figure~\ref{fig:VT} against those calculated from the
  modeled rotation curve;
\item the tracers velocity dispersions of figure~\ref{fig:dispersions} against the dispersion
  profiles calculated from $V(r)$;
\item the circular velocities of the Maser regions;
\item $V_\odot$ as determined above from maser observations.
\end{itemize}
Denoting with $x_i$ all these different quantities (evaluated at their radii) and with $x_{i,exp} $
their observational constraint, a standard $\chi^2=\sum_{i=1}^{N} [x_i-x_{i,exp}]^2/\delta
x_{i,exp}^2$ is employed, and with our data, $N=40$.

The complete set of parameters is \{$\rho_H$, $R_H$, $M_B$, $M_D$, $R_D$, $R_\odot$\} which specify
the modeled galaxy rotation curve, plus the anisotropies \{$\b_1$, $\b_2$\} of the two populations
which determine the dispersion velocity profile. In practice however, the number of relevant
parameters can be reduced. In fact,
\begin{itemize}
\item As discussed in section~\ref{sec:dispersions}, we use $\b_1-\b_2=0.5$, and study $\b_2$ in the
  range $\b_2=-0.5\pm0.5$;

\item $R_D$ is constrained by observations, as discussed in section~\ref{sec:components}; moreover
  the fits strongly prefer large values of $R_D$ (especially in the NFW case) forcing it to be at
  the boundary of the allowed values;

\item $M_B$ and $M_D$ are free parameters that lie between a minimum and a maximum value. As we will
  see below the fits prefer often the lowest values for them so the minimum is on the boundary, and
  they play effectively a minor role in the minimization;

\item In view of the quite precise knowledge of the sun's angular velocity, $V_\odot/R_\odot\simeq
  30.3\pm0.3\,\km/\se\,\kpc$ as recalled above, we express $R_\odot$ in terms of $V_\odot$ by means
  of the above relation.
\end{itemize}
Finally, through the mass modeling there is a one-to-one correspondence between $\r_H$ and the
circular rotation at the sun $V_\odot=V(r_\odot)$ (all the other parameters kept fixed), and it is
technically convenient to trade $\r_H$ for $V_\odot$, which is used in its place as independent
variable.

\begin{figure*}[t]
\centerline{\includegraphics[width=1\columnwidth,height=.4\columnwidth]{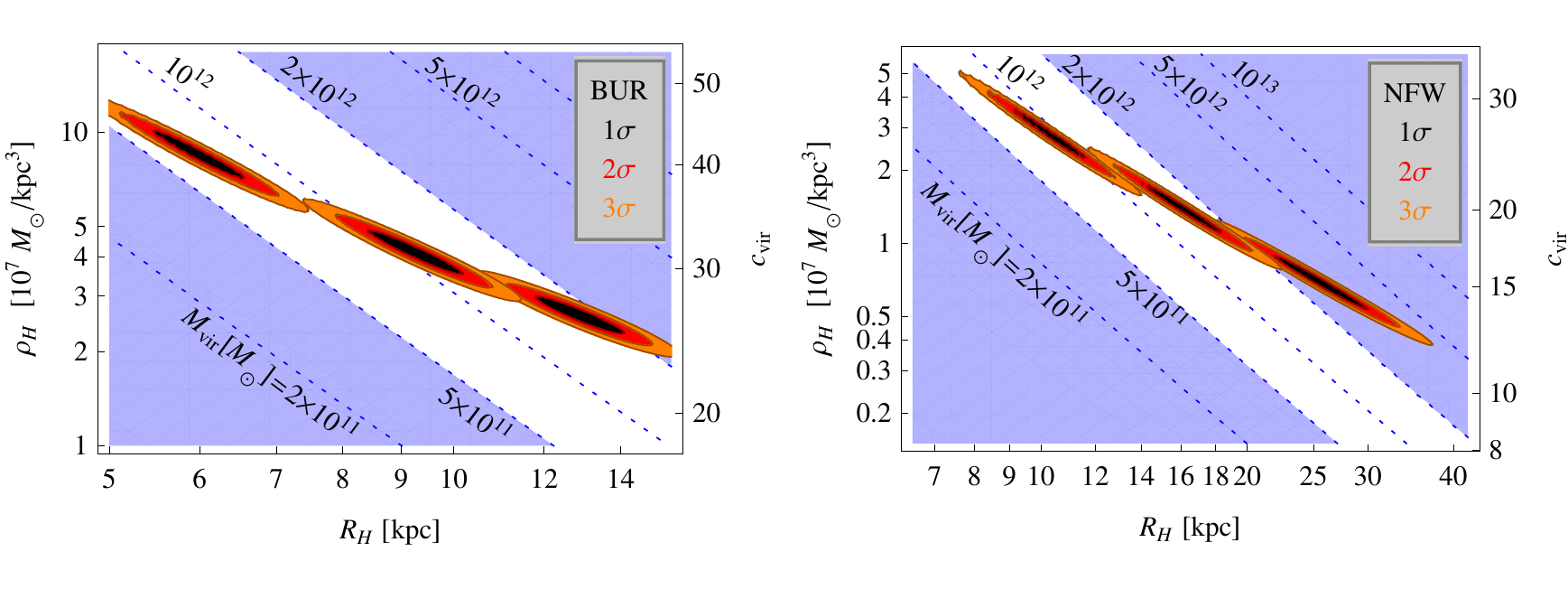}}
\vspace*{-3ex}
\caption{Best fits and confidence regions in the $\rho_H$--$R_H$ plane at 1, 2, 3$\s$ C.L.  (black,
  red, orange, yellow, two-parameter contours) for the Burkert (left) and NFW (right) profiles.  In
  each frame the three regions correspond to tracers dispersion anisotropy $\b_0=0,-0.5,-1$ (left to
  right).  The superimposed blue (solid and dotted) contours mark $M_{vir} [10^{12}\,\Ms]=0.1$--$10$
  and the shaded blue region is disfavored by estimates of the total MW mass.  The thick green
  dotted line shows the $c_{vir}$-$M_{vir}$ relation as resulting from cosmological
  simulations. Parameters other than $R_H$ and $\r_H$ are fixed at their best fit preferred values
  (see text).\label{fig:rhofits}}
\end{figure*}

\subsection{Galaxy mass modeling results}

\begin{figure*}[t]
\centerline{\includegraphics[width=.93\columnwidth]{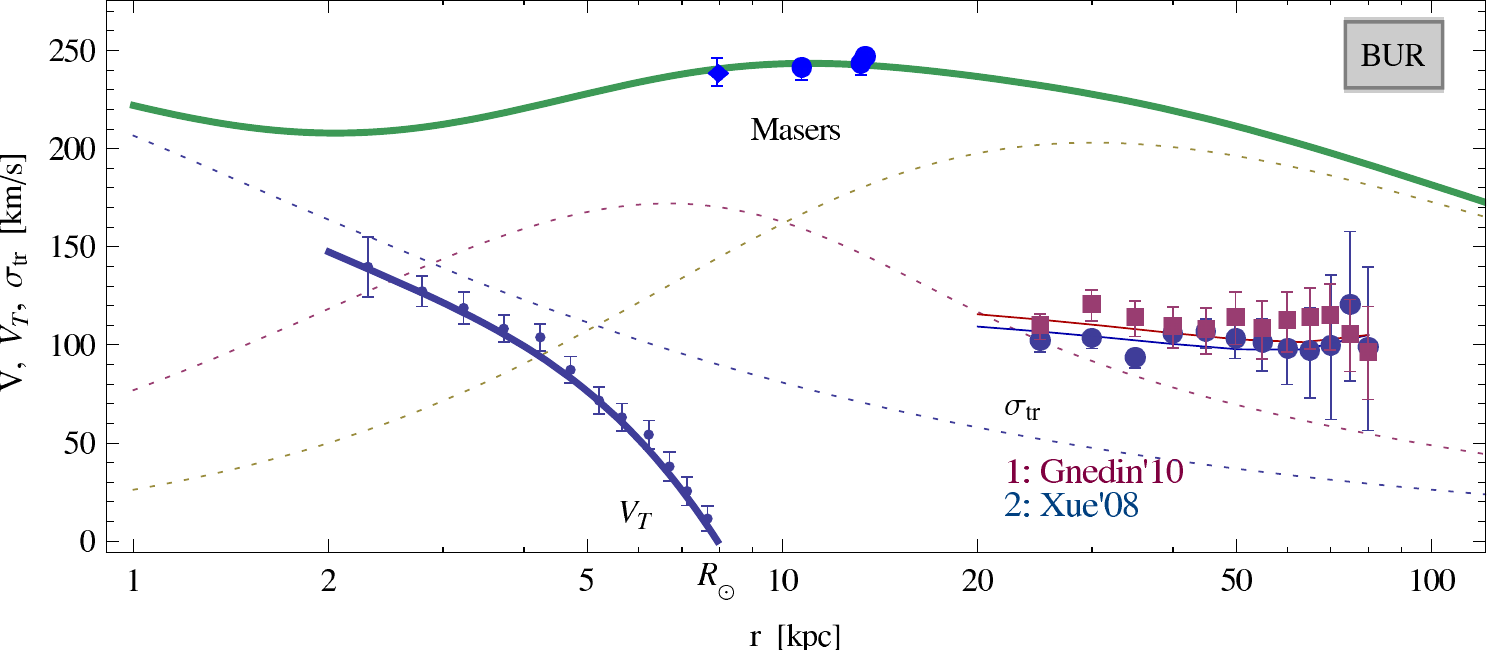}}
\bigskip\medskip
\centerline{\includegraphics[width=.93\columnwidth]{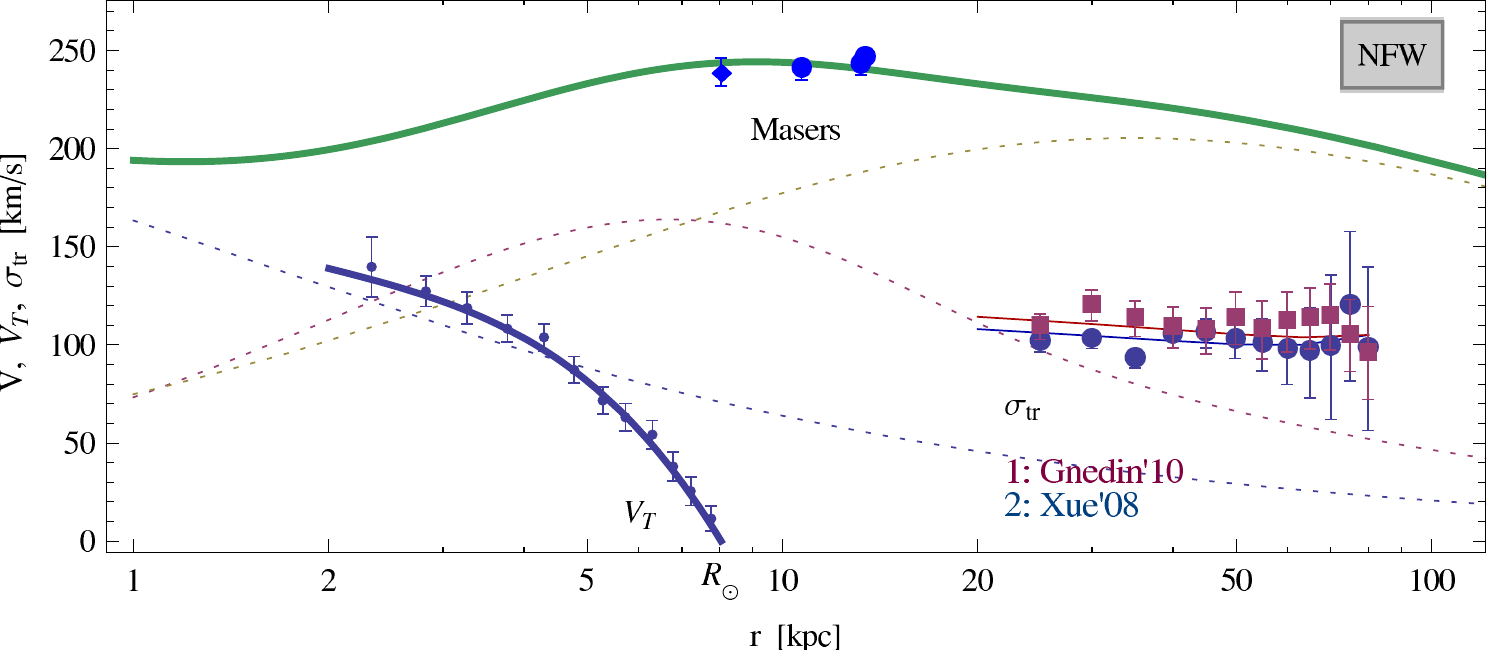}}
\caption{Best fit rotation curve, terminal velocities and velocity dispersions for the URC Burkert
  profile (upper panel) and for the NFW profile (lower panel), compared with the adopted
  observational data on terminal velocities, maser data and halo tracers velocity dispersions (left
  to right).  Dotted curves show the bulge, disk and DM halo separate contributions (left to
  right)\label{fig:bestfits}}
\end{figure*}

\begin{figure*}[t]
\centerline{\includegraphics[width=.82\textwidth]{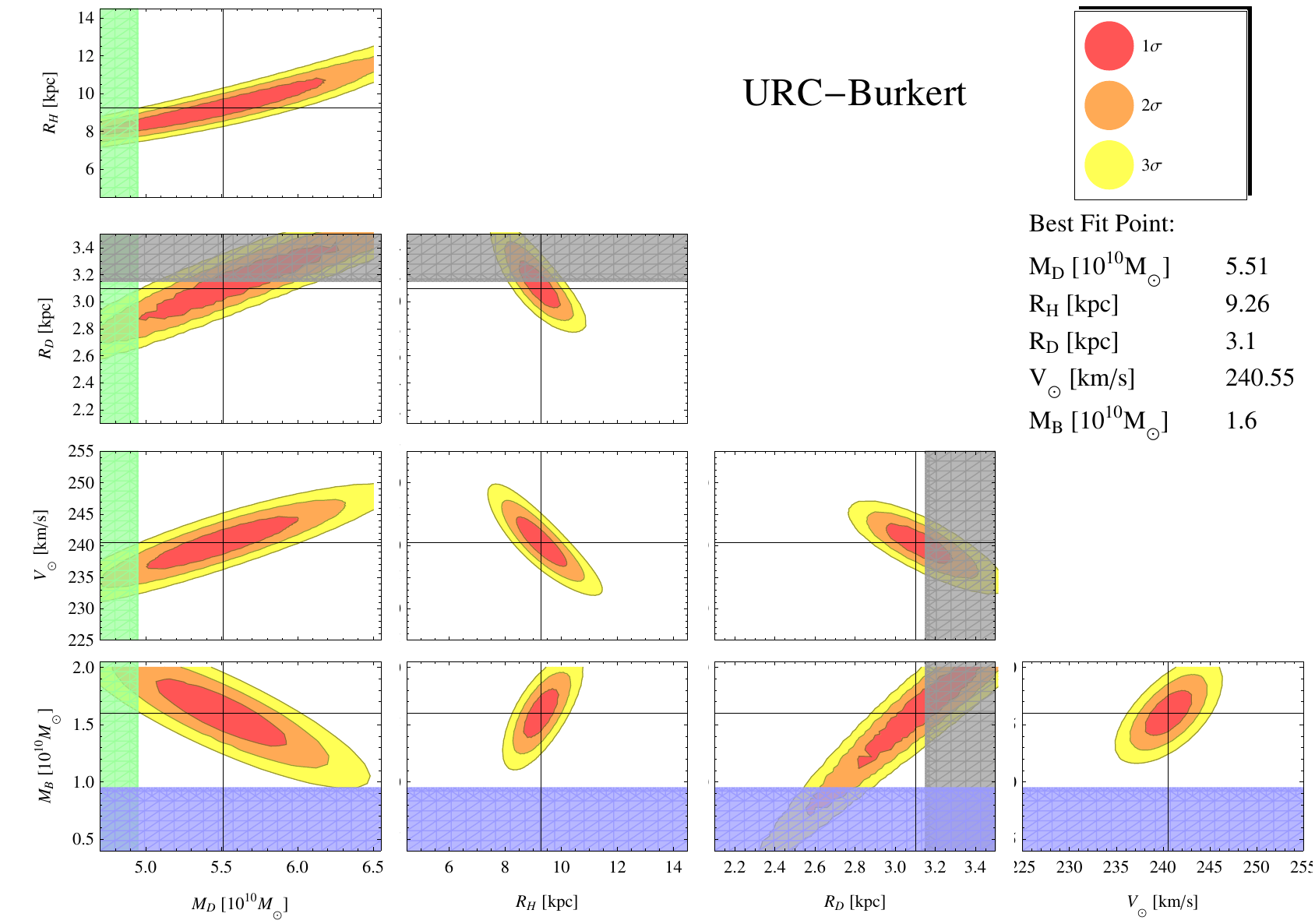}}
\centerline{\includegraphics[width=.82\textwidth]{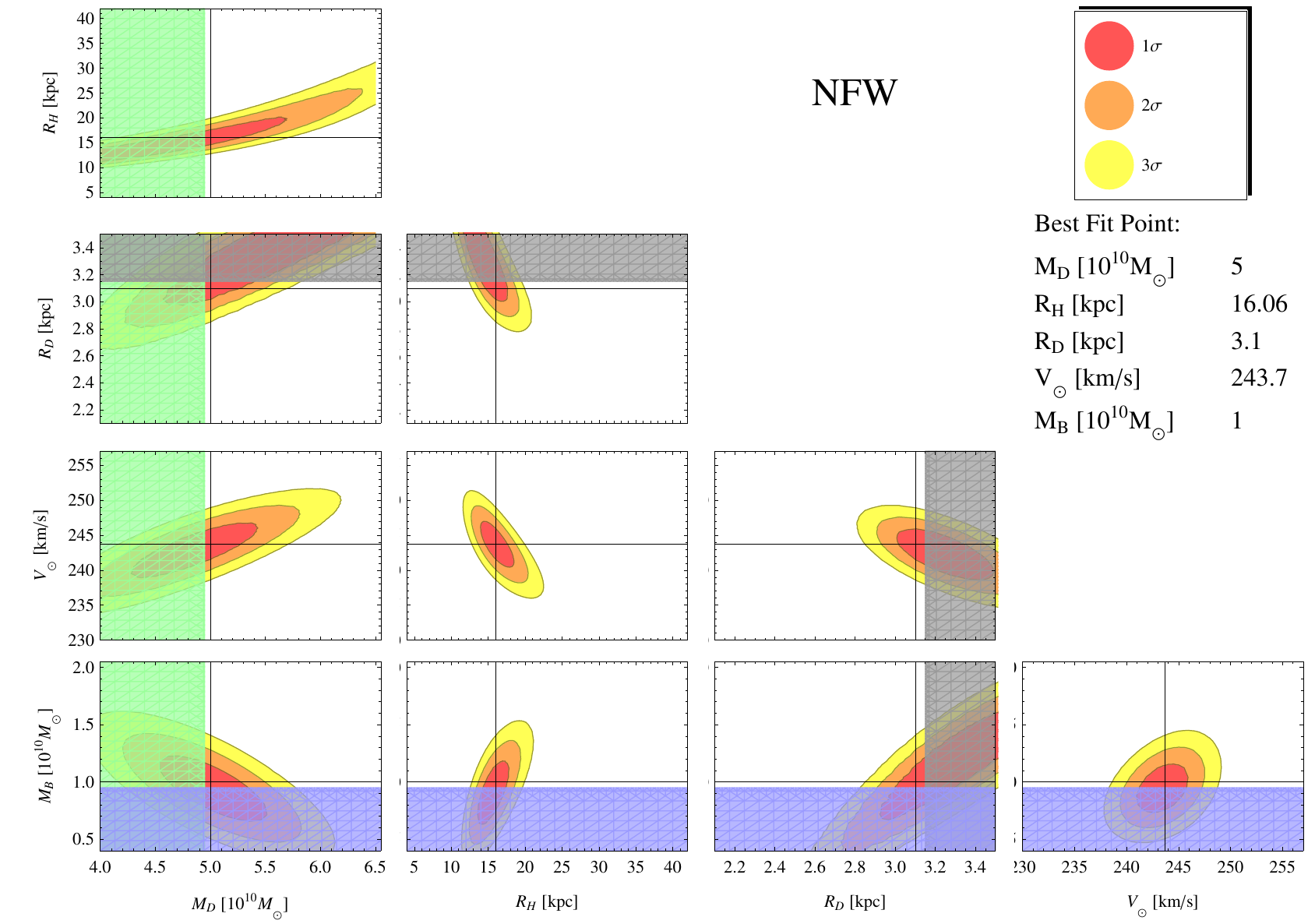}}
\vspace*{-1ex}
\caption{Two dimensional sections of $\chi^2$ for the URC Burkert (upper) and NFW (lower) profiles,
  passing through the best fits, with fixed $\b_2=-0.5$. The shadowed regions mark the values of
  $M_B$ (green left vertical band), $M_D$ (blue bottom horizontal band) and $R_D$ (the other grey
  bands), disfavored by bulge and disk studies.  The contours mark $\Delta\chi^2=(2.3, 6.18,11.8)$,
  i.e.\ the \emph{two-parameter} confidence levels of (68.27\%, 95.45\%,99.73\%).
  \label{fig:corr}}
\end{figure*}

The fits have been performed with values of $\b_2$ in a range of $\b_2\in[-1, 0]$, and we report in
figure~\ref{fig:rhofits} the resulting best fit halo parameters and confidence intervals for each
case of $\b_2=0$, $-0.5$, $-1$ (left to right in each frame).  In figure~\ref{fig:bestfits} we show
the central (preferred) best fit Galactic rotation curves, and in figure~\ref{fig:corr} we show the
sections of $\chi^2$, describing also the effect of the other astrophysical parameters and their
correlations.

As discussed above, given the estimate of $\b_2\simeq -0.5$ in \cite{Kafle:2012az}, we consider this
as our preferred fit point, and the other two as extremal cases.  With these three choices, the
values of the best reduced $\chi^2_{DOF=40}$ are respectively 0.59, 0.41, 0.35 for the Burkert
profile, and 0.9, 0.46, 0.35 for the NFW profile.  The fit is further improved with even more
negative values of $\b$, a fact which hints to possible tension between data and the model. Thus a
careful reconsideration of the uncertainties in the tracers velocity dispersions could lead to a
softening of this problem. Nevertheless, due to the low values of $\chi^2_{DOF}<1$, we can say that
both models are able to provide reasonable fits to the data.

From figure~\ref{fig:rhofits} we note that, as anticipated, the predicted values of $R_H$ are
strongly dependent on the values of the dispersion anisotropy $\b_2$. The possible variation of the
latter is thus the main source of uncertainty, driving the variation in $R_H$ and the ranges
presented in table~\ref{tab:best} for the halo parameters and derived quantities.  The observational
(or theoretical) likelihood profile of $\b_2$ is not known at present and thus one can not give a
definite evaluation of confidence intervals by marginalizing on this dominant uncertainty.
Nevertheless, the adopted range in $\b_2$ allows us to estimate the expected ranges of parameters,
in a consistent picture fitting all observational constraints, by extending the 95\% C.L.\ fit
uncertainties in a flat range of $\b_2$.

In figure~\ref{fig:bestfits} we display the best fit rotation curves, where the fitted terminal
velocities, velocity dispersions, and maser circular velocities are compared with data.  In
figure~\ref{fig:corr} we display the cross correlations between the fit parameters $M_D$, $M_B$,
$R_D$, $R_H$, $V_\odot$, by plotting the sections of $\chi^2$ through the best fit point, as well as
the constraints on astrophysical parameters. As a result of these constraints the best fits are
sometimes constrained to lie just at the boundary of these regions. This is true in particular for
$R_D$, and for $M_B$, $M_D$ in the NFW case.  In table~\ref{tab:best} we report the DM halo
parameters according to the preferred fit and also quote parameter uncertainties.

\begin{table}[t]
\begin{center}$
\renewcommand\arraystretch{1.65}
\begin{array}{|cc|c|c|}
\hline
    & &\qquad \text{BUR} \qquad& \qquad\text{NFW}\qquad \\
\hline
 \rho _H &~ [10^7M_{\odot }/\text{kpc}^3]~ &     4.13_{-1.6}^{+6.2} &   1.40_{-0.93}^{+2.9} \\
 R_H & \text{[kpc]} & 9.26_{-4.2}^{+5.6} &   16.1_{-7.8}^{+17.} \\
 V_{\odot } & \text{[km/s]} &  241._{-7.4}^{+11.} & 244._{-7.7}^{+6.3}   \\
\hline
\hline
 R_{\odot } & \text{[kpc]} & 7.94_{-0.24}^{+0.36}  &   8.08_{-0.2}^{+0.2} \\
 \rho _{\odot } & [\text{GeV/}\text{cm}^3] &   0.487_{-0.088}^{+0.075} &   0.471_{-0.061}^{+0.048} \\
 ~~M_{r<50\, \text{kpc}} & [10^{12}M_{\odot }] &   0.45_{-0.20}^{+0.35} & 0.48_{-0.15}^{+0.20} \\
 ~~M_{r<100\, \text{kpc}} & [10^{12}M_{\odot }] &     0.67_{-0.33}^{+0.67} &   0.81_{-0.32}^{+0.60} \\
 M_{\text{vir}} & [10^{12}M_{\odot }] &    1.11_{-0.61}^{+1.6} &   1.53_{-0.77}^{+2.3} \\
 c_{\text{vir}} & \text{[$\Delta $=200]} &   31.4_{-5.3}^{+13.} & 20.1_{-7.1}^{+11.}\\
\hline
\end{array}$
\end{center}
\caption{Best fit halo parameters and uncertainties for the Burkert and NFW profiles.  
  The best fit values are relative to $\b_1=-0.5$, while the reported 
  ranges correspond to 2\,$\sigma$ intervals (95.45\% C.L.)  calculated in the 
  \emph{two-parameter} space $V_\odot$-$R_H$ (see figure~\ref{fig:corr}) by considering 
  also the range $\b_1=[0,-1]$.\label{tab:best} As a result, quoted uncertainties should 
  not be considered as statistical gaussian intervals, but only as estimates of expected values.}
\vspace*{1ex}
\end{table}

From the parameter variations in table~\ref{tab:best} and from figures~\ref{fig:rhofits} and
\ref{fig:corr}, we see that the Milky Way mass modeling is still quite uncertain, and this is mainly
due to our ignorance of the dominant baryonic mass components (bulge and disk) and even more of the
kinematics in the outer galaxy.  The latter may be expected to improve in the near future due to
more extensive stellar surveys and maser probes; on the other hand, the knowledge of the mass
enclosed in the central region appears to be essentially hampered by the severe extinction at the
relevant wavelengths.  Still, some messages emerge from the fit.

The best fit Burkert model features a total halo mass of $M_{vir}\sim10^{12}\,\Ms$, a central DM
density of $\sim4\times10^7\,\Ms/\kpc^3$ and a core radius of $\sim9\,\kpc$.  The emerging mass
model results very similar to that of the URC relative at spirals like the Galaxy, although in the
two cases the gravitational potential is obtained in completely different ways.  The variations of
$\b_2$ can significantly affect the mass modeling, and drive the total mass also to values larger
than $2\times 10^{12}\,\Ms$.  The Burkert best fit values in table~\ref{tab:best} give $\mathop{\rm
  Log}(\r_HR_H/\Ms\pc^2)\simeq 2$, in perfect accordance with the trend obtained for other galaxies
in a wide span of magnitudes~\cite{donato09}. This is also the correlation between $\r_H$ and $R_H$
which can be inferred from the left figure~\ref{fig:rhofits}, so that this relation is stable under
variations of $\beta_2$.

We also analysed the model featuring a NFW DM profile plus baryons, shown in the right panel of
figure~\ref{fig:rhofits}, as well as in the lower panels of figure~\ref{fig:bestfits}
and~\ref{fig:corr}. In figure~\ref{fig:rhofits} we also show the contours of the total MW virial
mass $M_{vir}$ and the concentration parameter $c_{vir}$ on the right axis (using an overdensity
parameter $\Delta_{vir}=200$).

As one can see the allowed range of halo scale radii turns out to be quite wide,
$R_H=10$-$30\,\kpc$. In addition for NFW, unlike the Burkert case, the correlation between $\r_H$
and $R_H$ in the right panel of figure~\ref{fig:rhofits}, is of the form $\r_H R_H^{2}\sim$const.,
which reflects just the velocity constraints along the curve (both maser velocities and halo
dispersions), in practice just a byproduct of the parametrization of the DM density through $\r_H$,
devoid of deep significance.  Similarly for the $V_\odot$-$R_H$ correlation in the central frames of
lower figure~\ref{fig:corr}.

In addition, two more remarks can be made. First, the NFW plus baryons best fit models require
minimal bulge and disk masses as well as the largest disk scale radius, all at the boundary of the
constraining intervals.  This can be clearly appreciated in figure~\ref{fig:corr}, where these
parameters are varied and ther correlation shown.  If one fixes the more reasonable value of
$R_D\simeq 2.5\,\kpc$, the NFW fit worsens to $\chi^2_{DOF}\simeq2.2$, corresponding to a
probability of less than $10^{-4}$ (the Burkert fit instead raises to $\chi^2_{DOF}=0.7$ which is
still very good).  Thus, the acceptable fits of the NFW plus baryons model are only achieved by
stretching the astrophysical parameters to their allowed values.

Second, a fairly high value for $c_{vir}\simeq 20\pm 10$ is obtained for the NFW profile.  Current
$\Lambda$CDM cosmological N-Body simulations imply the relation $c_{vir}= 9.6
M_{vir}/(h^{-1}10^{12}M_\odot)$~\cite{Prada:2011jf} (with $\Delta_{vir}=200$), which is also
depicted on figure~\ref{fig:rhofits} for comparison.  The Galaxy reasonably has
$M_{vir}\gtrsim10^{12} M_\odot$, so according to simulations we should have found $c_{vir}<10$, a
value significantly lower than the best fit one.  If we impose the above $c_{vir}$-mass
relationship, the resulting NFW Milky-Way mass model becomes an unacceptable representation of data.
In table~\ref{tab:best} we also report the predicted value of $c_{vir}\sim 31^{+13}_{-5.3}$ for the
Burkert fit; we stress that, because in the URC scenario $c_{vir}$ is just a structural parameter,
so far unconstrained by galaxy formation scenarios, this value should not be compared with
$\Lambda$CDM simulations which only produce cusped DM profiles, and instead should be taken as the
outcome (prediction) of our fitted MW model.

It is important to stress that the uncertainty in the tracers anisotropy discussed above can not be
invoked to release this tension of $\Lambda$CDM simulations, since a) it would require very negative
values such as $\beta_2\sim -3$, i.e.\ the existence of very tangential motions at odds with results
of cosmological simulations, and b) the total MW mass would end up being unacceptably high, $\sim
10^{13}\,\Ms$. Considering that the NFW best fits are already obtained by stretching astrophysical
parameters to their extrema, models that solve the high concentration issue would also be too
contrived.

For these reasons, we can conclude that present data tend to prefer a cored profile with respect to
a cusped one.  Part of this preference can be tracked down to the terminal velocities, which require
the lowest possible mass enclosed in the inner region ($r<8\,\kpc$), given that the disk and bulge
already provide enough gravitational potential (see figure~\ref{fig:corr}).\footnote{In fact, we
  performed a similar analysis for the Einasto profile, finding even worse fits in the inner
  region.}  Acceptable fits can still be achieved, but at the price of finding a solution in tension
with the ones arising within the $\Lambda$CDM paradigm.

\medskip 

The value of the DM density at the Sun's radius $\rho_\odot$ is also an outcome of the fit, and it
turns out to be consistent with the profile independent determination derived in~\cite{nostro} (see
discussion below).

Recalling that the main source of uncertainty in our fit comes from the poor knowledge of the outer
tracers anisotropies $\b_i$ and the new recent knowledge on the latter, it is useful to compare our
findings with the results of some other works.

In \cite{cu09}, MW mass models were built in the process of estimating of DM local density. They
have considered both NFW and URC Burkert dark matter halos and tested them with a number of
dynamical observables for the Galaxy (similar but not coincidental with those in the present paper,
in particular with a different likelihood function).  Both models have been found to agree with
their observational data, with values of parameters somewhat similar but coincidental with ours.
Among those an higher value of $c_{vir}$. In \cite{Catena:2011kv}, with a similar global fit, they
study also the DM phase space distribution. In view of the uncertainties in the halo tracers
dynamics and also in the other available observative constraints, we consider quite premature such
an analysis.
 
By assuming state-of-the-art models for the distribution of baryons in the
Galaxy,~\cite{Iocco:2011jz} used microlensing and dynamical observations of the Galaxy to show that
these data are in good agreement with the predictions of $\Lambda$CDM Dark Matter profiles.
                                                                                                                                                                                
In \cite{Deason:2012ky} 2000 distant Blue Horizontal Branch stars at Galactocentric distances of
$15\,\kpc < r < 50\,\kpc$ were used as kinematic tracers of the MilkyWay mass distribution. Their
density was modeled as an oblate, power-law halo embedded within the spherical power-law DM
potential. Be means of distribution function method they obtained the power-law potential exponent
and the velocity anisotropy of the halo tracers. The resulting outer circular velocity profile for
the Milky Way when reproduced by a NFW halo, led to a high concentration value ($c_{vir} = 20$) in
agreement with present results.

In \cite{McMillan:2011wd} a simple method for fitting a MW mass model to observational constraints
was presented (not too different from those we used in this paper). By means of a Bayesian approach
they took input from observational data and expectations from the self consistent NFW plus disc
model.  They also forced the value of $c_{vir}$ to lie in a range compatible with N-Body
simulations, i.e.\ between $7$ and $12$.  For the values of the disc scale lengths of $3.00 \pm
0.22\,\kpc$ Solar galactocentric distance of $8.29 \pm 0.16\,\kpc$ and a circular speed at the Sun
of $239 \pm 5 \km/\se$; MW stellar mass of $6.43 \pm0.63\times10^{10}\,\Ms$; a virial mass of $1.26
\pm 0.24\times 10^{12}\,\Ms$ and a local dark matter density of $0.40 \pm 0.04\,\GeV/\cm^{3}$ their
model fits observations.
                                                                                                                                                                                
It is also worth to stress that the Deason, McMillan or Iocco approaches are not much different from
ours, but they did not consider cored DM profiles. These works, with the present one, support the
view that in the MW a NFW plus baryon model can reproduce data, but with somewhat stretched values
of the astrophysical parameters.


\clearpage

\begin{figure}[t]
  \centerline{\includegraphics[width=.65\columnwidth]{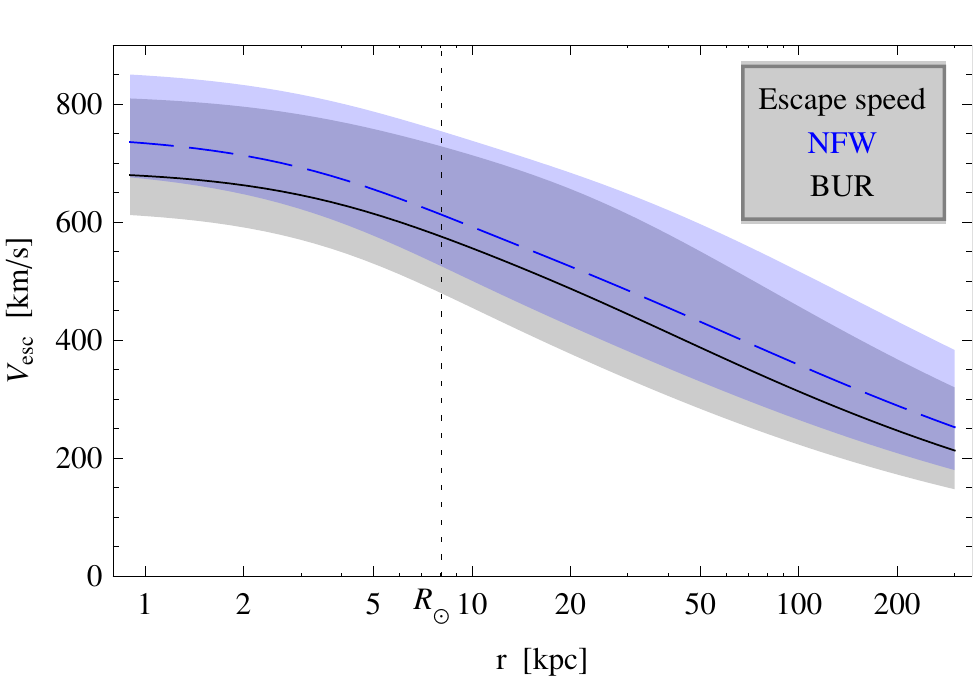}}
  \caption{Galactic escape velocity as a function of galactocentric radius $r$ for the best fit
    URC Burkert (black continuous) and NFW (blue dashed) models, with their 95\% uncertainties in the
    halo parameters, including the uncertainty in the anisotropy $\b_2$.  The dotted line marks the
    Sun's radius.\label{fig:escape}}
\vspace*{3ex}
\end{figure}

\section{Direct DM search: local density and escape velocity}
\label{sec:direct}

The derived mass models provide us with the local value of DM density.  For the Burkert profile one
finds $\rho_\odot\simeq 0.49^{+0.08}_{-0.09}\,\GeV/\cm^3$, for the NFW values of $\simeq
0.47^{+0.05}_{-0.06}\,\GeV/\cm^3$.  These values are clearly consistent with the profile independent
estimate of~\cite{nostro}, as it must be since the gauss law and formula (11) there are satisfied.
Nevertheless, the present global MW mass modeling has helped in reducing the uncertainties, from 0.2
to roughly $0.1\,\GeV/\cm^3$, because the global fit constrains the values of $M_D$ , $R_\odot/R_D$,
$V_\odot$ and rotation-curve slope more strictly than the set of the upper-lower limits taken
in~\cite{nostro}. The central value is also slightly higher, due to the increased rotation speed at
the Sun $V_\odot$, which is now better known.

From the mass models one can derive the Galactic escape velocity, whose value at the Sun's location
is important for direct searches through DM nuclear recoil, because it affects the DM velocity
distribution.  We plot in figure~\ref{fig:escape} the escape velocity as a function of radius, out
to 300\,\kpc. One can observe that at the Sun's location the Burkert and NFW best fits lead to quite
similar estimates of the escape velocity, we have $V_{esc}^{BUR}(R_\odot)\simeq576\pm124\,$km/s, and
$V_{esc}^{NFW}(R_\odot)\simeq613\pm114\,$km/s. These values are quite large and reflect the
uncertainty due to the tracers anisotropy $\b_2$. They are however in agreement with recent
estimates of the local escape speed using RAVE data~\cite{Smith:2006ym}.  Considering the overall
variation of escape speed among the two models, one finds a large uncertainty of $\sim250\,$km/s,
i.e.\ roughly 50\%.  This should be taken into account in direct search experiments, especially in
the regime of light ``WIMP'' DM mass of the order of $\sim10\,\GeV$ or less, where one relies on the
upper tail of the velocity distribution to estimate the number of expected recoil events (see e.g.\
the analysis in~\cite{mccabe},~\cite{Catena:2011kv}).

The uncertainty in the escape velocity at large radii $r>25\,\kpc$ is also considerable, (i.e.\ from
300 to 600\,$\km/\se$ at $50\,\kpc$) and this has an impact when estimating the halo dispersion
velocities, through the selection of outliers versus acceptable halo tracers.  Clearly in the future
a modeling with this feedback would give a more self-consistent picture (see e.g.~\cite{brown09}
for such an attempt).

\begin{figure}[t]
\centerline{\includegraphics[width=.65\columnwidth]{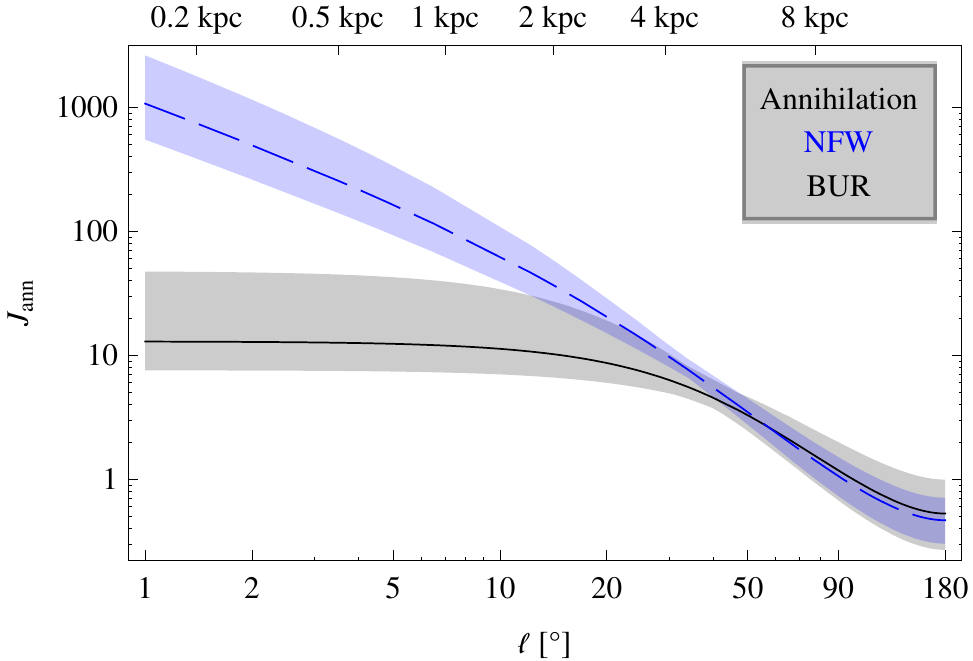}}
\caption{The ``prompt'' emission factor from DM annihilation as a function of the Galactic
  longitude, for the best fit Burkert (black continuous) and NFW (blue dashed) models, with their
  2$\sigma$ regions (95.45\% C.L.).  Note that for gamma rays in for instance the Fermi detector,
  below $<0.1\text{--}1^\circ$ the point-spread function would smear the observed profile, making it
  effectively cored for any DM profile.\label{fig:ann}}
\end{figure}

\section{Indirect DM search: annihilation}
\label{sec:ann}

The flux from DM annihilation is conveniently expressed in terms of the ``prompt'' emission factor
\be
J_{ann}(\ell)=\frac1{\bar\r_{\odot}^2\bar R_\odot}\int_{l.o.s.}\r_H^2({\mathbf x})\,{\rm d}{\mathbf x}\,,
\ee
which we normalize by using $\bar\r_\odot=0.4\,\GeV$ and $\bar R_\odot=8.3\,\kpc$.  The factor
$J_{ann}$,  as a function of the longitude $\ell$ from the galactic center, traces directly the
angular profile of the dominant observed flux from annihilation into gamma rays. For annihilation
into other (charged) particles, which are then subject to bremsstrahlung, scattering with ISR, and
nontrivial galactic diffusion, see~\cite{Cirelli:2010xx}.

In fig~\ref{fig:ann} we plot $J_{ann}$ for the URC  Burkert + baryons and for the NFW + baryons models.
We see that for each mass model the uncertainties in the galactic parameters lead to variations of
the expected flux of a factor of $\sim5$, in the innermost region.  On the other hand, in direction
of about 40--60$^\circ$ from the galactic center the flux is predicted within a factor of 2 only,
and independently of the profile chosen.

In fact, the differences between the two different mass models emerge only at $\ell<15^\circ$, and
with a clear discriminating power only below $\ell<10^\circ$, corresponding to $r<1.5\kpc$, i.e.\
inside the bulge region. One should thus bear in mind that this plot extrapolates the DM density
profile in the very central region where observations can not constrain it.  Indeed, the DM density
in the bulge region or at shorter scales may well deviate from purely cored or NFW profile, still
without modifying the present global fits.

For instance, if one is willing to consider the scenario in which a DM core results from baryon
feedback (mainly supernovae explosions) which erases the density cusp during galaxy formation, the
same mechanism may well leave a `mini-cusp' in the central region, which would give a small
contribution to the total mass inside the solar circle, but depending on the very inner density
slope, may contribute to an evident annihilation signal from the inner zone, with a very localized
source region of at most few degrees in angular size.

\section{Conclusions}
\label{sec:conclusions}

In this paper we have derived a model for the DM in the Milky Way, by adopting state of the art
inner terminal velocities, probes of maser star forming regions, and outer stellar halo velocity
dispersions.  We performed our analysis for both a cored (Burkert) and cusped (NFW) profiles.  

Our fit confirms that the presence of a dark component in the Galaxy is beyond any doubt.  We
derived the preferred ranges of DM halo parameters $\rho_H$, $R_H$ and their correlation in
figures~\ref{fig:rhofits}, and reported the values and derived parameters in table~\ref{tab:best}.
The results show that the DM halo is still subject to a considerable degree of uncertainty. This is
due to the limited constraining power of the available observations, especially in the outer stellar
halo region, and in the Galactic center, bulge region. The main source of uncertainty has been
traced to the unknown velocity dispersion anisotropy of the tracers in the stellar halo, which is
still very poorly constrained by observations, and which drives the uncertainty in the halo scale
radius $R_H$.  Nevertheless, thanks to recent more precise measurements of the rotation speed at the
Sun's position $V_\odot$ via maser emission in high star-forming regions, the fitted halo parameters
allow to draw some interesting conclusions.

First, if one considers a cored profile like the adopted URC-Burkert one, very good fits are
obtained, which can accommodate all available observations. The uncertainty in the tracers velocity
anisotropy is still driving the DM halo parameters uncertainty, resulting in the ranges reported in
table~\ref{tab:best}, with $R_H=9.26^{+5.6}_{-4.2}\,\kpc$ and $\r_H=4.13^{+6.2}_{-1.6}$ with quite a
strict correlation of $\r_HR_H\sim const$, even varying the tracers anisotropy $\beta_2$.  The value
of $\r_HR_H\simeq10^2$ is in excellent agreement with Dark-Matter fits in many other similar
external galaxies, and constitute a reference point on which one may base further analyses of the
Dark Matter in our Galaxy.

\medskip

We studied also the case of NFW profile and showed that, while it also can be used to fit the
observations, this was only possible by stretching other astrophysical parameters to their allowed
limits; e.g., the disk scale length had to be set to $3.1\,\kpc$ which is at its marginal upper
limit, and the bulge and disk masses had to be as small as possible. This is due to the profile of
inner terminal velocities, which require as low mass as possible in the central region, where
instead the NFW cusp is. Indeed, a similar analysis for the Einasto profile shows a further
worsening of this tension.  In other words, the present data rather prefer a cored DM distribution,
though the presence of standard NFW halo cannot be excluded and, if one decides to adopt it, a great
care must be taken in choosing the values of the model parameters.

Also, an important evidence is that especially due to the high $V_\odot$ value, the NFW halo turns
out to have a high $c_{vir}\sim20$, at odds with results from numerical simulations in the
$\Lambda$CDM setup.  Of course this might be due to the effect of dirty baryonic physics.  Recent
simulations have started to account for the effect of baryons which have a twofold effect, namely,
to: i) adiabatically contract the halo, increasing the inner halo slope even beyond 1--2, and ii)
expel the inner dark matter due to supernovae feedback. It is not clear whether the net final effect
is to reduce the concentration $c_{vir}$ even further with respect to the baryon-less case.  In this
scenario a DM profile much different from the original NFW one is likely to emerge and, in our
opinion, to converge with the URC one.

\medskip

In the light of the fitted halo parameters, we analyzed the predicted Galactic escape velocity
profile and the flux from DM prompt annihilation.  While consistent with estimates with different
methods, the escape speed is found to be very uncertain, independently from the profile chosen: at
the Sun's radius it ranges in the interval $500$--$750\,\km/\se$, which has an impact on direct dark
matter searches, especially for low (WIMP) DM masses of the order of $10\,\GeV$.  The escape speed
turns out to be quite uncertain also at large radii in the outer galaxy, with an impact on the
removal of unbounded outliers from the samples of stellar halo tracers.

Regarding the flux from DM annihilation, we assessed the discriminating power on the cored versus
cusped DM halo hypotheses of indirect DM searches toward the galactic center.  In
figure~\ref{fig:ann} we displayed the expected fluxes from prompt DM annihilation, showing that a
discrimination is possible only in the region inside $10^\circ$ (or $1.5\,\kpc$) from the Galactic
center. This is right in the bulge region, and may be subject to a considerable astrophysical
background uncertainty. Let us also stress in this respect that the observational data are able to
constrain the DM profile only at radii larger than about $2\,\kpc$; but because the physics leading
to the inner DM density profile is still not understood, it may well be the case that the DM profile
deviates from the simple core or cusp.

The DM density at the Sun's location is also an outcome of the mass modeling, and the values found
are $\r_H(R_\odot)\simeq0.49^{+0.08}_{−0.09}$ for the Burkert profile (and $0.471^{+0.05}_{−0.06}$
for NFW).  They are clearly in agreement with the profile-independent determination~\cite{nostro},
but its uncertainties are reduced by a factor of two from $\sim 0.2$ down to roughly $\lesssim
0.1\,\GeV/\cm^3$.  This is due to the fact that in the present global modeling the rotation curve is
more constrained by inner and outer galactic observations and the sun's rotation speed is known
better.

In the present study, we tried to assess realistic ranges for the DM halo parameters, but still
being far from claiming to give precise statistical significance intervals. The reason is due to the
lack of knowledge of realistic likelihood profiles for many important quantities (the stellar halo
anisotropy parameters but also the disk and bulge masses) so that any arbitrary choice of ad-hoc
priors would translate in unrealistic or unstable results.  We try to avoid such unjustified
predictions in parameters, both as a matter of principles and also to avoid complicating the
development of the field of Galactic modeling as well as the one of dark matter detection.

\section*{Acknowledgments} 
P.S.\ thanks the Gran Sasso Center for Astroparticle Physics (CFA) and INFN, Laboratori Nazionali
del Gran Sasso (LNGS) for hospitality during the completion of this work.

\bigskip

\begin{appendix}

\section{Terminal velocities}
\label{app:terminali}

\begin{figure}[h]
\centerline{\includegraphics[width=.55\columnwidth]{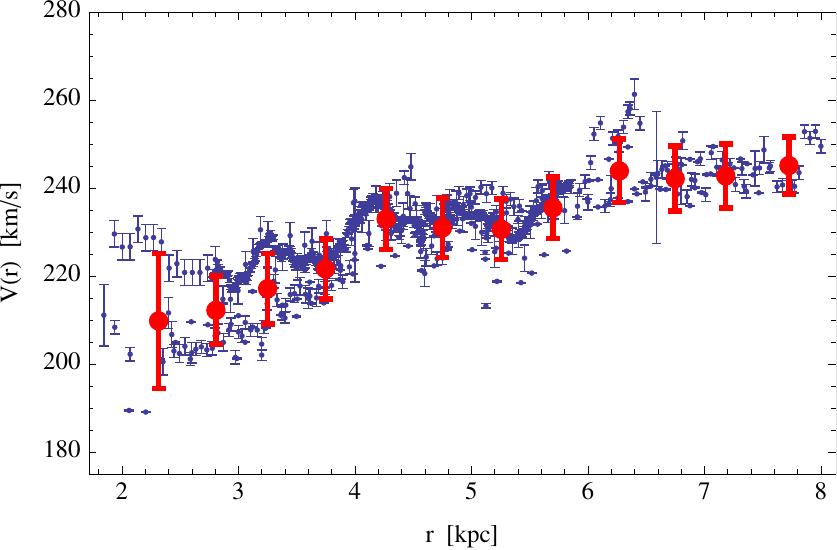}}%
\vspace*{-1ex}
\caption{Circular velocities reconstructed from terminal velocities (blue, thin) and values obtained
  after their binning (red, thick). For $V_\odot=242\,$\km/s and
  $R_\odot=8\,\kpc$.\label{fig:terminalibinnati}}
\end{figure}

We describe here the adopted data relative to terminal velocities in the inner Galactic region. We
use available data from~\cite{malhotra95,clemens85,alvarez90,McClure07} versus the absolute value
of the longitude.  Since the data show systematic variations due to local gas motion, it is
important not to trust either the small individual errors (often smaller than 1\,km/s) or the high
number of data points. Indeed, even an infinite increase in the number of measures would only amount
to a more precise map of the gas motion, and not imply a more precise determination of the circular
velocity. As a result, the adopted uncertainty in the circular velocities should follow the
dispersion of the data, which directly traces the local variations and does not become precise with
an increase of data points.  We perform a binning of $\sim0.5\,\kpc$ and consider the weighted
average and the dispersion inside each bin as data for the global fit.  The binned average is
performed by first binning each dataset separately (to avoid dominance of more populated datasets)
using original errors enlarged by 7\,km/s (to avoid the dominance of sets with tiny experimental
error). Then, inside each bin, we average the results from different datasets, and consider the
weighted average and the dispersion inside each bin as data for the global fit. The resulting binned
terminal velocities are shown in figure~\ref{fig:VT}.  In figure~\ref{fig:terminalibinnati} we show
also the original data, together with the binned velocities, where for convenience we have
reconstructed the circular velocities for $V_\odot=242\,$km/s and $R_\odot=8\,\kpc$.

\end{appendix}

\end{document}